\documentclass[12pt,preprint]{aastex}

\begin{document}
\shorttitle{Minivoids in the Local Volume}

\shortauthors{A. Tikhonov}

\title{Voids in the SDSS Galaxy Survey}
\author{Anton V.\ Tikhonov }
\affil{St.Petersburg State University, Saint-Petersburg, Russia}
\email{ti@hotbox.ru, avt@gtn.ru}

\begin{abstract}

Using the method of searching for arbitrary shaped voids in the
distribution of volume-limited samples of galaxies from the DR5
SDSS survey, we have identified voids and investigated their
characteristics and the change in these characteristics with
decreasing $M_{lim}$ (from -19.7 to -21.2, $H_0 = 100$~km/s/Mpc)
--- the upper limit on the absolute magnitude of the galaxies
involved in the construction of voids. The total volume of the 50
largest voids increases with decreasing $M_{lim}$ with a break
near M* = -20.44 --- the characteristic value of the luminosity
function for SDSS galaxies. The mean density contrast in voids
increases with decreasing $M_{lim}$ also with a weak break near M*
The exponent of the dependence of the volume of a void on its rank
increases significantly with decreasing $M_{lim}$ starting from
$M_{lim} \sim -20.4$ in the characteristic range of volumes, which
reflects the tendency for greater clustering of brighter galaxies.
The averaged profile of the galaxy density contrast in voids has a
similar pattern almost at all $M_{lim}$. The galaxies mostly tend
to concentrate toward the void boundaries and to avoid the central
void regions; the density contrast profile is flat in the
intermediate range of distances from the void boundaries. The
axial ratios of the ellipsoids equivalent to the voids are, on
average, retained with changing $M_{lim}$ and correspond to
elongated and nonoblate void shapes, but some of the voids can
change their shape significantly. The directions of the greatest
void elongations change chaotically and are distributed randomly
at a given $M_{lim}$. The void centers show correlations
reflecting the correlations of the galaxy distribution on scales
$(35 - 70)h^{-1}$~Mpc. The galaxy distribution in the identified
voids is nonrandom --- groups and filaments can be identified. We
have compared the properties of the galaxies in voids (in our
case, the voids are determined by the galaxies with absolute
magnitudes $M_{abs} < M_{lim} = -20.44$, except for the isolated
galaxies) and galaxies in structures identified using the minimum
spanning tree. A bimodal color distribution of the galaxies in
voids has been obtained. A noticeable difference is observed in
the mean color indices and star formation rates per unit stellar
mass of the galaxies in dense regions (structures) -- as expected,
the galaxies in voids are, on average, bluer and have higher
$log(SFR/M_{star})$. These tendencies become stronger toward the
central void regions.
\end{abstract}

\vspace{1ex}

\section{Introduction}

The distribution of galaxies is a complex "cosmic network". The
walls, filaments, and voids observed at the present epoch reflect
both linear and nonlinear evolution of clustering. The pattern of
the observed clustering out to $20-25h^{-1}$~Mpc can be described,
for example, by a power law with a complex dependence of the
exponent on the luminosity, color, and other properties of
galaxies with the subsequent transition to homogeneous
distribution, with the structures being traceable on scales
exceeding the scale of homogeneity (Tikhonov 2006a, 2006b). The
nature of such clustering depends on many small- and large-scale
factors such as the cosmological parameters, the environments of
galaxies and clusters, their formation history, the distribution
of dark matter, and the scenario according to which the luminous
and dark matter are related and evolve. The characteristics of
voids have long been considered as tests of cosmological models.
Regoes and Geller (1991) found that in their model for the
formation of structures, certain initial conditions lead to the
formation of a "cellular" structure with voids similar to those
observed in galaxy surveys. Voids are the forming components of
the large-scale structure. In recent years, various authors have
considered in detail both observational and theoretical aspects of
the existence and evolution of voids detected both in galaxy
catalogs and in the dark matter halo distributions obtained in the
$\Lambda CDM$ model calculations of the N-body problem by Hoyle
and Vogeley (2004), Gottlober et al. (2003), Shandarin et al.
(2004), Croton et al. (2004), Benson et al. (2003), Colberg et al.
(2005), and Patiri et al. (2006b). The void statistics are closely
related to the methods of calculating the galaxy clustering; for
example, VPF (Void Probability Function) provides information
about the high-order correlation functions (Croton et al. 2004).

One important problem of the modern theory of the formation of
structures is that according the $\Lambda CDM$ model for the
evolution of dark matter structures including the $\Lambda$-term
attributable to the existence of "dark energy," much matter must
be present in voids, while the expected number of galaxies in
voids is not observed (Peebles 2001).

Sheth and van de Weygaert (2004) developed a model for the
distribution of void sizes and evolution in terms of the
hierarchical clustering scenario. Furlanetto and Piran (2006)
developed an analytical model that predicted the shape of the
distribution of void sizes. In particular, they found that because
of the so-called "bias" effect, the voids in the galaxy
distribution are considerably larger than those in the dark matter
distribution.

The shapes of voids are of interest along with the spectrum of
their sizes. Based on numerical calculations, Ike (1984) concluded
that, in most cases, the voids between filaments must be nearly
spherical in shape. Plionis and Basilakos (2002) analyzed the
distribution of void sizes and shapes in the PSCz survey and
compared them with the artificial distributions obtained in terms
of various CDM models. Shandarin et al. (2004) found, in
particular, that their large voids defined as regions with a
density lower than a given value in the smoothed density field of
the dark matter distribution are essentially nonspherical.

Patiri et al. (2006a) found that the distribution
of galaxies in voids in the distribution of 2dFGRS
galaxies differs significantly from a random one.

It has been firmly established that, compared to the general
distribution, the galaxies in voids have bluer colors, lower
luminosities, and higher star formation rates. In addition, a
higher abundance of disk galaxies is observed among the galaxies
in voids (Peebles 2001; see also Patiri et al. 2006b). Rojas et
al. (2004, 2005) confirmed these tendencies by analyzing the
photometric and spectroscopic properties of the galaxies in voids.
Hogg et al. (2004) considered the dependence of the galaxy color
and luminosity distributions on the density contrast and found
that, on the one hand, the most luminous galaxies populate the
densest regions, while the blue galaxies are present mostly in
low-density regions, and, on the other hand, the mean parameters
of the distributions in absolute magnitude and color change only
slightly with over- density. Based on SDSS galaxies, Baldry et al.
(2004) showed that the galaxy color distribution is bimodal and is
described well by two Gaussians. They obtained a fitto the (u-r)
--- color–absolute magnitude relation and compared the luminosity
functions for red and blue galaxies. Hoyle et al. (2005) found
significant differences between the luminosity functions of the
galaxies in voids and dense regions.

Patiri et al. (2006b), who analyzed the voids in the DR4 SDSS
survey, found no significant differences between the mean
parameters of the field and void galaxies. In this paper, we use a
different approach to selecting a "check sample" of galaxies for
comparison with the properties of the galaxies in voids --- we
selected the galaxies of the "check sample" in high density
structures located entirely outside the void boundaries.

In this paper, we also analyze the variations of void
parameters with luminosity and perform a correlation
analysis of the distribution of void centers.

\section{THE DATA}

The spectroscopic redshifts are expected to be obtained for about
106 galaxies and 105 quasars within the framework of the Sloan
Digital Sky Survey (SDSS) based on photometric data for a sky
region $10^4$ square degrees in area in the Northern Galactic
Hemisphere in five bands ($u$, $g$, $r$, $i$, $z$) with a limiting
magnitude of $r = 22.5$ (York et al. 2000; Stoughton et al. 2002)
once the program has been fully implemented. The photometric data
were used for a homogeneous selection of various classes of
objects to obtain their spectra. Two types of galaxies were chosen
for determining the redshifts from the list of objects classified
as extended ones: galaxies with a Petrosian magnitude $r < 17.77$
and a surface brightness exceeding 24 m/$\square$". formed the
Main Galaxy Sample (the number of objects in the final SDSS
version is $\sim900000$); the LRG (Luminous Red Galaxies) list
includes galaxies with very red colors and $r < 19.5$ (the number
of objects in the final SDSS version is $\sim100000$). In this
paper, we analyzed data from the fifth data release DR5 SDSS
(www.sdss.org, Adelman-McCarthy et al. 2006).

When analyzing the DR5 data, we selected a rectangular part from
the region of spectroscopic sky coverage for the convenience of
allowance for the boundary conditions in determining the void
boundaries and for ensuring sample completeness. In the ($\lambda,
\eta $) coordinate system of the survey, the selected region is
$-48^\circ < \lambda < 48^\circ$, $6^\circ < \eta < 36.5^\circ$.

The Main Galaxy Sample is an apparent-magnitude-limited survey,
which determines the method of constructing the volume-limited
sample to eliminate incompleteness in radial coordinate --— we set
the limit on the $r$-band absolute magnitude for the sample
galaxies equal to $M^0_{lim} = r_{lim}-25-5log(R_{max}
(1+z_{max}))-K(z)$,where $r_{lim} = 17.77$ was taken as the
limiting $r$-band magnitude, $K(z)$ is the K-correction, and
$R_{max}$ is the chosen far boundary in radial coordinate
corresponding to zmax.The u and r magnitudes used here were
corrected for extinction.

To estimate the absolute magnitudes of the galaxies, we used a
mean Ê correction for SDSS galaxies in the form $K(z)=
2.3537z^2+0.5735z-0.18437$ (Hickage et al. 2005; see also Blanton
et al. 2003). The metric distances were recalculated from the
redshifts with the Hubble parameter $H_0 =
100$~km~s$^{-1}$Mpc$^{-1}$, $h=H/H_0$, where $H$ is the true value
of the Hubble constant, and the density parameters
$\Omega_{\Lambda} = 0.7$, $\Omega_0 = 0.3$ (see, e.g., Hogg 1999).

\section{THE METHOD}

The void construction algorithm presented here has already been
applied by Tikhonov (2006b) and Tikhonov and Karachentsev (2006)
and is basically similar to the algorithm described by El-Ad and
Piran (1997). The voids were constructed in the distribution of
"bright" galaxies with absolute magnitudes $M_{abs}$ (in the $r$
band) lower than a certain value of $M_{lim}$. Here, we searched
for voids containing a certain number of galaxies from the
volume-limited sample with $M_{abs} < M_{lim}$. For the
void-forming galaxies, we determined the mean distance to the
nearest neighbor $R_{n}$ and the standard deviation $\sigma_n$.If
there was no neighbor with $M_{abs} < M_{lim}$ in the sphere of
radius $R_{n}+\sigma_n$ around a particular galaxy of this sample,
then this galaxy was excluded from the list of galaxies involved
in the construction of voids. Thus, these excluded galaxies could
fall into voids. The mean distance from the isolated galaxies to
the nearest neighbor is considerably larger than that for isolated
pairs (when two galaxies lie in the sphere of radius
$R_{n}+\sigma_n$) --- it is close to the mean distance between the
galaxies for pairs. Thus, having eliminated the influence of
isolated galaxies, we obtain more "stable" voids.

Next, we successively searched for galaxy-free seed spheres inside
the sample volume (first, the largest sphere is searched for) and
then expanded them by adding spheres whose centers are inside the
already fixed part of the void and whose radii $R_{sph}$ are not
smaller than the radius of the seed sphere multiplied by the
coefficient $k = 0.9$ ($R_{sph} > 0.9 \cdot R_{seed}$, where
$R_{seed}$ is the radius of the seed sphere). The voids are
assumed to be located entirely within the geometrical boundaries
of the sample.

The voids constructed in this way (at $k = 0.9$) have arbitrary
shaped volumes. On the other hand, the voids are separated from
each other and fairly "thick" throughout the volume, which allows
them to be approximated by triaxial ellipsoids.

There exist other methods of searching for voids that are more
commonly used in analyzing artificial dark matter distributions
(see, e.g., Shandarin et al. 2006). In this approach, a smoothed
(e.g., with a Gaussian filter) density field is constructed (the
parameters of the resulting structures depend on the smoothing
length) and a certain threshold local density that separates the
low-and high-density regions is specified. In this case, the voids
can be highly irregular in shape. A brief overview of the methods
and references can be found, for example, in Tikhonov (2006b) and
Patiri et al. (2006a).

\section{THE DEPENDENCE OF VOID PROPERTIES
ON LUMINOSITY}

To analyze how the properties of the voids vary over a wide
luminosity range of the galaxies forming them, we chose the
redshifts limits $z_{min} = 0.02$ and $z_{max} = 0.1$ (sample A).
The upper limit on the absolute magnitude of the galaxies in the
volume-limited sample with these boundaries is $M^0_{lim} =
-19.67$. For this sample, the number of galaxies is $N = 47892$,
the mean density is $\rho \approx 7 \cdot 10^{-3}h^{3}$,
$R_{n}+\sigma_n \approx 3.0h^{-1}$~Mpc. The range of limits
$M_{lim}$ on the absolute magnitudes of the galaxies involved in
constructing the voids in which we analyzed the dependence is from
 $-19.7$ (40526 galaxies after the exclusion of isolated galaxies) to
$-21.2$ (1241 galaxies). We chose this range in such a way that
the sample was volume limited and that it contained a sufficient
number of galaxies. The resolution (the separation between the
grid points) was about 1.7 Mpc in all cases.

In the same vein, we analyzed sample B with the redshift limits
$z_{min} \approx 0.080$ and $z_{max} = 0.115$ chosen in such a way
that the sample covered the same volume (about $6.82 \cdot 10^6
\cdot h^{-3}$)~Mpc$^{3}$) as sample A. In this case, $M^0_{lim} =
-20.00$, $R_{n}+\sigma_n \approx 3.4h^{-1}$~Mpc.

\begin{figure}
\centerline{
\includegraphics[]{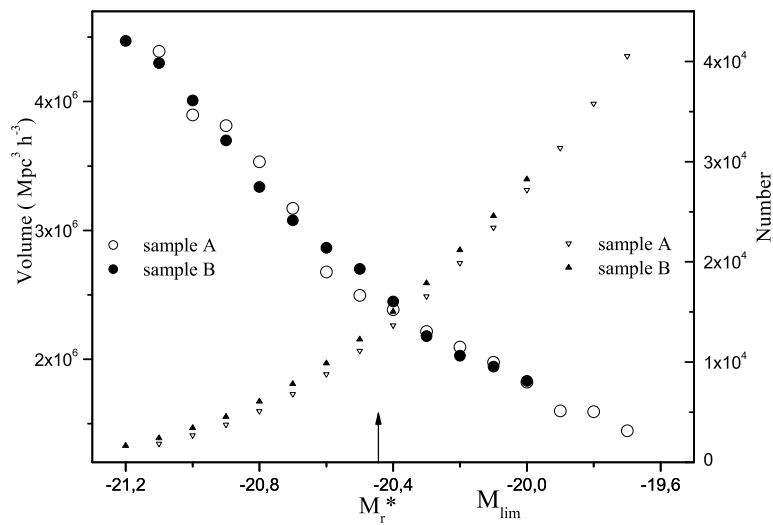}
} \figcaption{Total volume of the 50 largest voids as a function
of the limiting absolute magnitude $M_{lim}$ of the galaxies
involved in constructing the voids for samples A and B. Also shown
is the number of galaxies.}
\end{figure}

Figure 1 shows the total volume of the 50 largest voids in samples
A and B and the number of void- forming galaxies as a function of
$M_{lim}$. The increase in total volume occurs synchronously with
the decrease in $M_{lim}$ and in the number of galaxies with
$M_{abs}< M_{lim}$ (except for the isolated galaxies). A
significant break in the dependence is observed near $M_{lim} =
-20.5$, i.e., immediately after the characteristic value of $M^* =
-20.44$ of the luminosity function for SDSS galaxies followed by a
faster growth of the total volume. This is not just the result of
the corresponding decrease in the number of void-forming galaxies
(on the contrary, the decrease in the number of galaxies slows
down). The 50 largest voids occupy from about $21\%$ ($M_{lim} =
-19.7$) to $64\%$ ($M_{lim} = -21.1$) of the entire sample volume.

\begin{figure}
\centerline{
\includegraphics[]{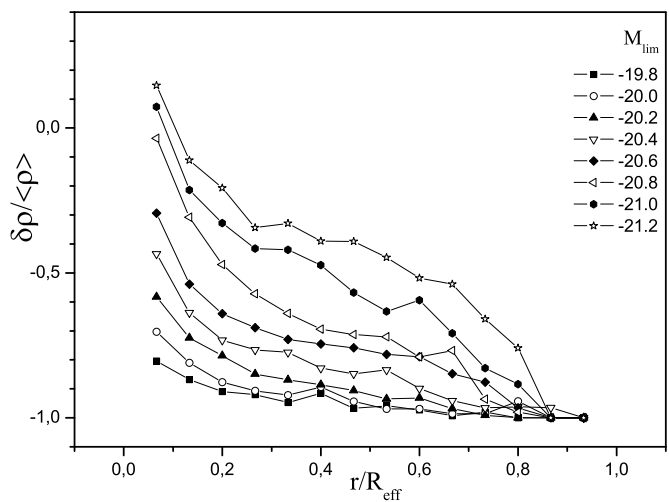}
} \figcaption{Mean overdensity of the galaxies in voids with
$R_{seed}>15$~Mpc with the distance $r$ from the void boundaries
normalized to $R_{eff}$ versus $M_{lim}$. }
\end{figure}

Figure 2 shows the dependence of the mean density contrast
(contrast profile) for the galaxies of sample A
$\delta\rho/<\rho_{VL}>$ (where $<\rho_{VL}>$ is the mean density
of the galaxies with $M_{abs}< -19.67$ in sample A) that fell into
a layer inside a particular void with the distance $r$ from the
void boundaries normalized to the effective void radius $R_{eff} =
(3 \cdot Vol/4\cdot\pi)^{-1/3}$ ($Vol$ stands for the void volume)
on the limiting absolute magnitude $M_{lim}$. We averaged the
density contrast profile inside all voids with $R_{seed}>15h^{-1}$
Mpc. The density contrast profile has common characteristic
features for different $M_{lim}$: the galaxies concentrate to the
void boundaries; there are virtually no galaxies in the central
regions ($r/R_{eff}> 0.7-0.8$); the density contrast profiles of
the galaxies in voids are flat up to $r/R_{eff}=0.3-0.2$ followed
by a significant increase in contrast. At $M_{lim} < -21.0$, the
galaxy density near the void boundaries (in the first bin) is
higher than the mean density ($\delta\rho/<\rho_{VL}> > 0$).

\begin{figure}
\centerline{
\includegraphics[]{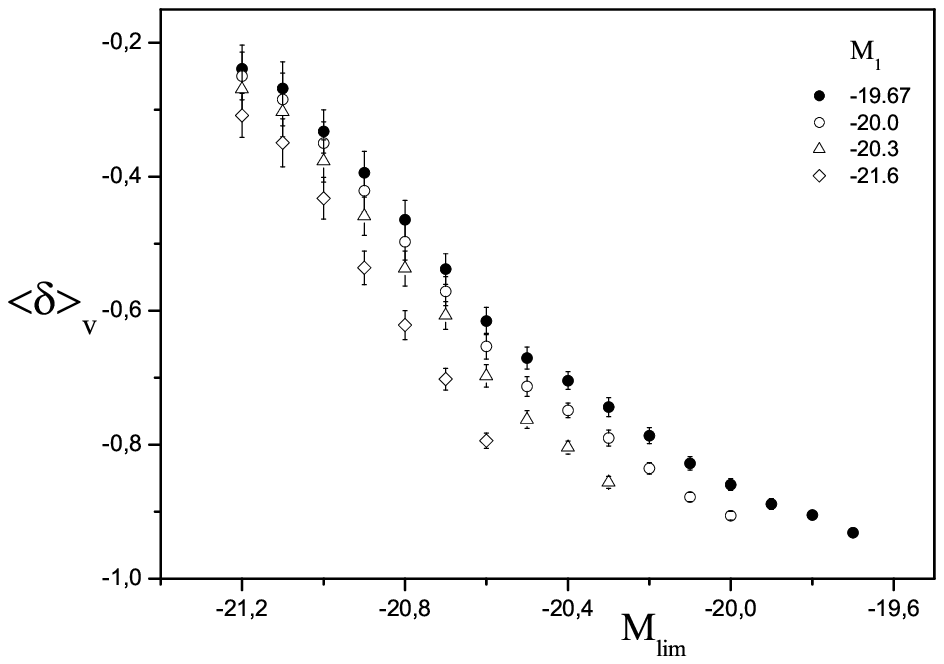}
} \figcaption{Mean overdensity $<\delta_v>$ of the galaxies in
voids (relative to the mean density of the sample galaxies with
$M_{abs} < M_1$) versus $M_{lim}$ for various $M_1$ (different
symbols are used for different values).}
\end{figure}

Figure 3 shows the dependence of the mean density contrast for the
galaxies in voids $<\delta_v>$ with $R_{seed}>15h^{-1}$ Mpc on
$M_{lim}$ for four limits $M_1$ on the absolute magnitude of the
galaxies retained in the volume-limited sample (from which the
mean density in the formula for the density contrast $<\delta_v> =
\delta\rho/\rho_{M_1}$) was obtained). In each of the four cases,
the change in $M_{lim}$ began from $M_{lim} = M_1$. An indistinct
break in the dependence can be distinguished at $M_{lim} \approx
-20.6$. At fixed $M_{lim}$, the mean density contrast is lower
(the voids become "emptier") at lower $M_1$. Convergence is
observed for all $M_1$ as $M_{lim}$ decreases. This indicates that
the most luminous galaxies form a stable skeleton of the structure
(the fraction of isolated galaxies decreases with decreasing
$M_1$). In all cases, the mean density contrast $<\delta_v> < 0$,
i.e., the identified voids are physically separated low-density
regions.

After compiling the list of voids (assigning the three-dimensional
grid points to a particular void), we determined the void centers
and calculated the moments of inertia of the bodies formed by the
voids. We analyzed the void shapes based on the parameters of the
equivalent ellipsoids. We constructed a $3\times3$ matrix of the
moments of inertia $I_{ij}$ and used the condition
$det(I_{ij}-\lambda \cdot E)=0$, where $E$ is a $3\times3$ unit
matrix, to find its eigenvalues $\lambda_i$, which are equal to
the principal moments of inertia, from which the semiaxes of the
equivalent ellipsoid were determined. The eigenvectors of matrix
$I_{ij}$ give the directions of the semiaxes. The direction of the
greatest void elongation coincides with that of the largest
semiaxis of the equivalent ellipsoid.

The void shapes were analyzed for the first 20 identified voids.
In general, the change in void configuration and shape with
decreasing $M_{lim}$ is indicative of an irregular change in the
configuration of the entire large-scale structure (in our
approach, the influence of isolated galaxies was eliminated). The
void centers are displaced significantly and the order of void
identification changes.

\begin{figure}
\centerline{
\includegraphics[]{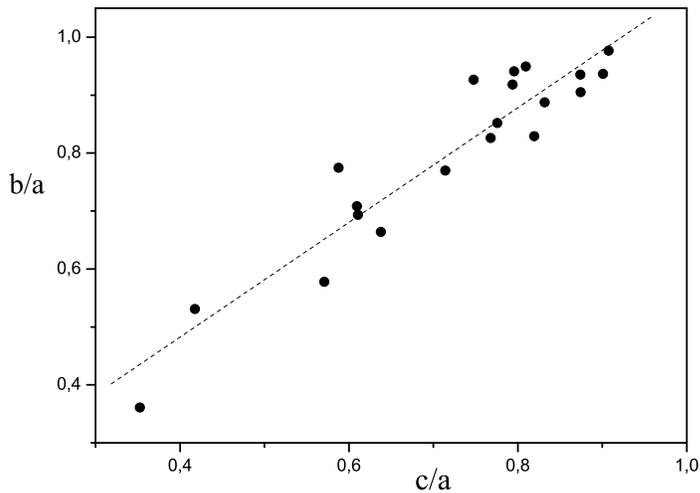}
} \figcaption{Example of the correlation between the smallest-
to-largest (c/a)and medium-to-largest (b/a)axial ratios of the
equivalent ellipsoids of the first 20 identified voids. The slope
of the linear fit is $\phi = 1.0$ and $M_{lim} = -20.3$.}
\end{figure}

\begin{figure}
\centerline{
\includegraphics[]{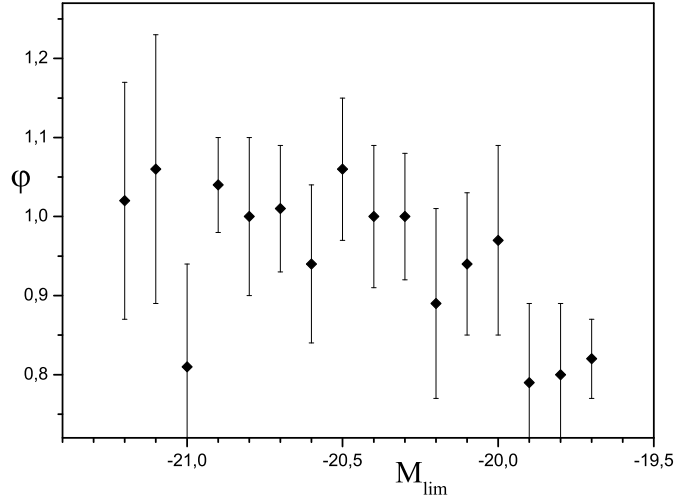}
} \figcaption{Slope $\phi$ of the linear fit to the distribution
of the b/a and c/a ratios (in the approximation of a triaxial
ellipsoid) for the first 20 (at given $M_{lim}$)identified voids
versus $M_{lim}$.}
\end{figure}

\begin{figure}
\centerline{
\includegraphics[]{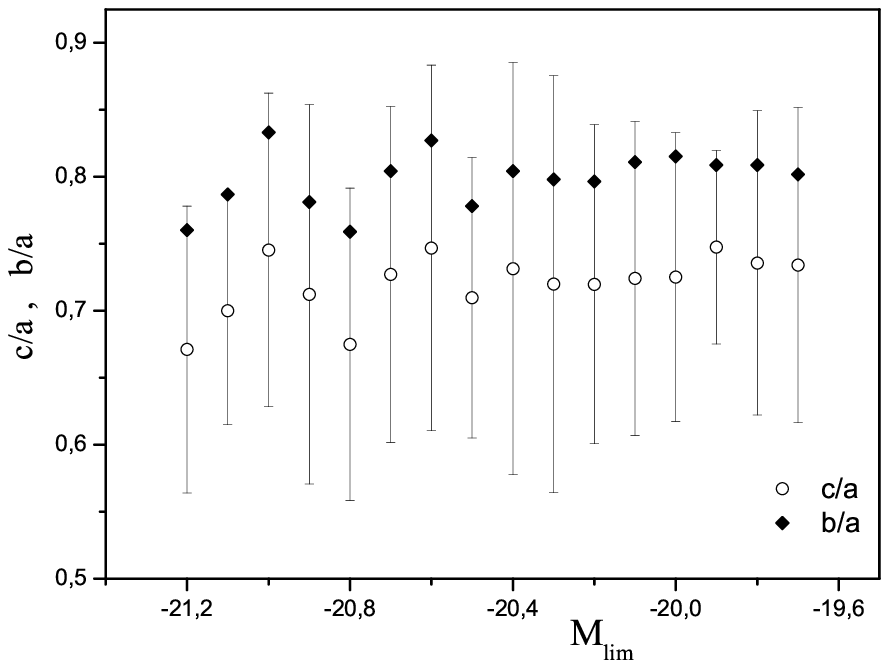}
} \figcaption{Mean c/a and b/a ratios for the first 20 voids
versus $M_{lim}$. The distribution errors $\sigma$ are given for
the c/a ratios.}
\end{figure}

The b/a and c/a axial ratios of the equivalent ellipsoid, where a,
b and c are the largest, medium, and smallest axes, respectively,
are correlated (see Fig.4 for $M_{lim}=-20.3$). Figure 5 shows the
mean slopes $\phi$ with the errors of the linear fits to the b/a
and c/a distributions for various $M_{lim}$. The dependence is
stable starting from $M_{lim} = -20.3$ and is approximately equal
to 1, i.e. the voids are predominantly in the shape of a slightly
elongated cucumber.

The mean c/a and b/a show a small trend at $M_{lim}< -20.6$ (for
the c/a ratio, Fig.6 presents $\sigma$ of the distribution in the
form of an error) --- the rations decrease with increasing
luminosity and the voids, on average, become more elongated. At
the same time, the relation between c/a and b/a is retained in the
entire $M_{lim}$ range --- the void oblateness is small and
changes only slightly. For all $M_{lim}$, the cases where c/a <
0.5 for any of the first 20 voids are rare (the void has a
significant elongation)—most of the voids have c/a > 0.6 (the
cases of "circular" voids with c/a > 0.9 are also rare).

\begin{figure}
\centerline{
\includegraphics[]{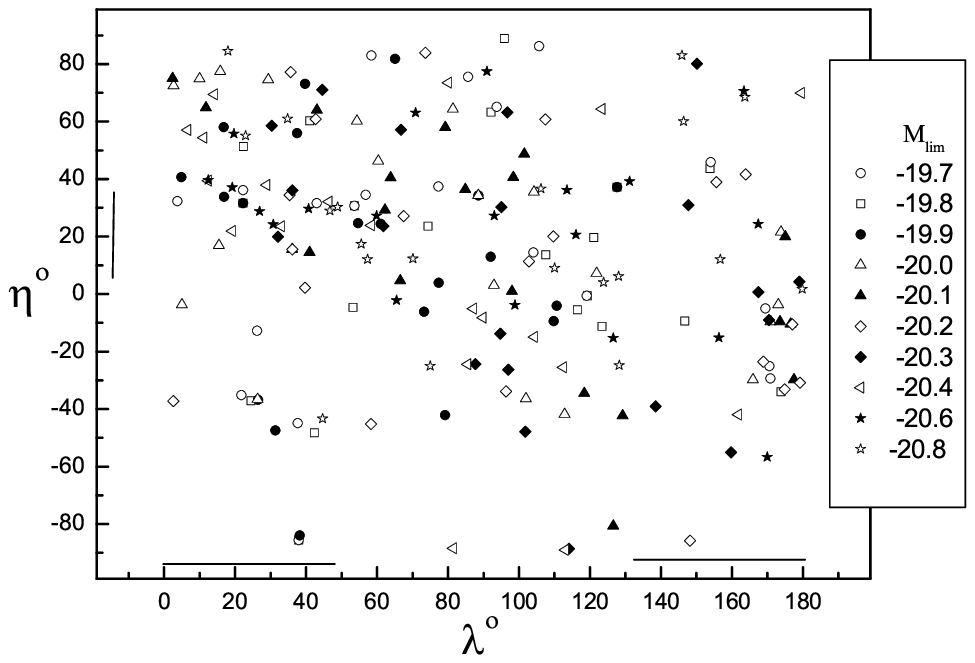}
} \figcaption{Directions of the greatest elongations (the largest
axes a of the equivalent ellipsoids) of the first 20 voids
identified in sample À for $M_{lim}$. The ranges of the directions
corresponding to the geometrical boundaries of the sample are
shown.}
\end{figure}

Figure 7 presents the directions of the greatest elongations for
the first 20 voids for the chosen $M_{lim}$ range. The void
orientations change significantly even for neighboring values of
$M_{lim}$ (although there are also cases where the void
orientation is retained in a certain $M_{lim}$ range, e.g., near
$\lambda = 25^o$, $\eta = -35^o$) and fill almost the entire area
of possible directions rather uniformly.

We also considered the variation in the parameters of a single
void that, for most values of $M_{lim}$, is the first void
identified by the algorithm and its shape is determined only by
the geometry of the distribution and does not depend on the
boundaries of other voids. The volume and $R_{seed}$ of this void
increase synchronously with decreasing $M_{lim}$ (Fig. 8); the
dependencies have a break at $M_{lim}=-20.6$ followed by a faster
increase in $R_{seed}$ and the volume. The orientation of this
void changes significantly and chaotically (Fig. 9). The c/a and
b/a ratios increase irregularly with decreasing $M_{lim}$ (Fig.
10) --- the void becomes "more circular" (except for
$M_{lim}=-21.2$ at which the surrounding structure apparently
changes greatly). The relation between c/a and b/a, along with the
direction of the greatest elongation, changes significantly and
irregularly.

\begin{figure}
\centerline{
\includegraphics[]{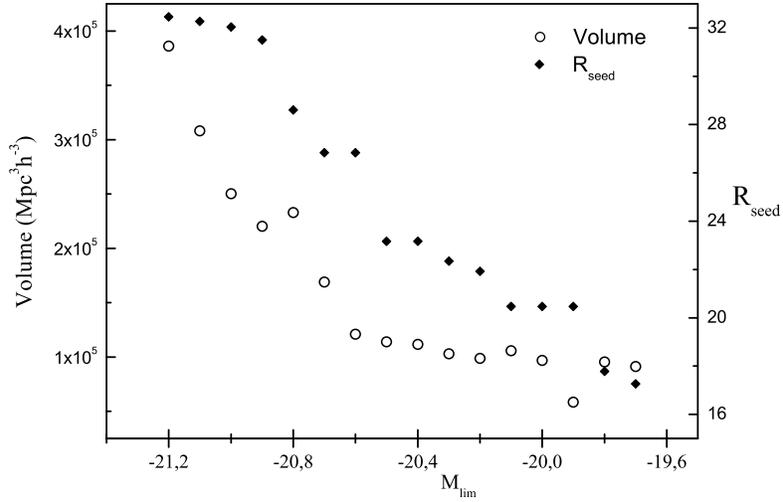}
} \figcaption{Volume and $R_{seed}$ of one first identified void
versus $M_{lim}$.}
\end{figure}

\begin{figure}
\centerline{
\includegraphics[]{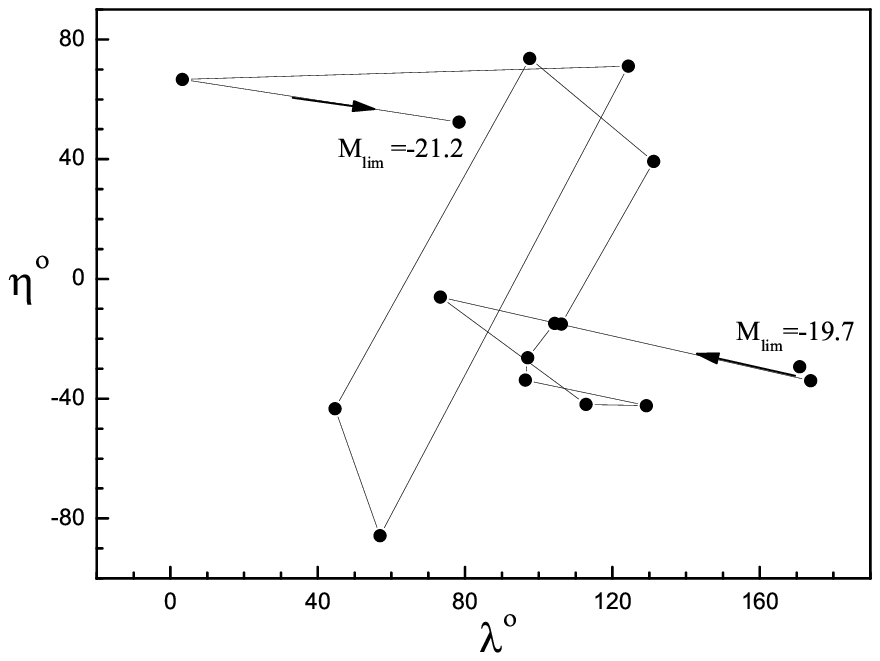}
} \figcaption{Evolution of the direction of the elongation of one
first void with $M_{lim}$.}
\end{figure}

The dependence of the void volume ($Vol$) on the void rank
($Rank$) at fixed $M_{lim}$ (the largest void has rank 1, the next
void has rank 2, etc.) may have a simple interpretation (Gaite and
Manrubia 2002). In particular, the break in this dependence
reflects the scale of the transition from a power-law galaxy
distribution to uniformity (Gaite 2005; Tikhonov 2006b). In this
paper, we analyzed the change in the slope $z$ of the $log(Vol)$
--- $log(Rank)$ relation with decreasing $M_{lim}$ in the chosen range
of volumes. We choose the range of volumes for each $M_{lim}$
between volume $V_1 \approx 10^4h^{-3}$Mpc$^3$ (the relation
exhibits a cutoff at the void ranks corresponding to volume V1,
which is probably determined by the constraint imposed on the
minimum $R_{seed}$ of the identified voids) and volume $V_2$
corresponding to the break in the $log(Vol)$ --- $log(Rank)$
relation, which is interpreted as the beginning of the transition
to uniformity in the galaxy distribution (Fig. 11). The power law
segment of the relation between $V_1$ and $V_2$ with exponent $z$
reflects a power-law galaxy distribution on small scales. The
exponent $z$ increases with decreasing $M_{lim}$ with a break near
$M^*$ (Fig. 12). Staring from $M_{lim} = -21.6$, the values of $z$
become larger than unity, which allows the formal fractal
dimension $D_z = 3/z$ of the galaxy distribution at given
$M_{lim}$ to be estimated (this dimension is considered as a
measure of the extent to which the sample volume is filled with
galaxies). The significant increase in $z$ with decreasing
$M_{lim}$ after the break reflects the well-known fact that more
luminous galaxies are clustered more strongly (see, e.g., Tikhonov
2006a). The values of $z < 1$ and the nearly flat pattern of
variation in $z$ to $M_{lim} = -21.4$ need a further study and an
explanation. The effective radius of a void with volume $V_2$,
$R_{eff} = (3 \cdot V_{2}/4\cdot\pi)^{1/3}$, can be associated
with the scale of the transition to uniformity at given $M_{lim}$.
In the range of values under study, this scale ($R_{eff}$) lines
in the range $19-27 \cdot h^{-1}$~Mpc and increases irregularly
with luminosity.

\begin{figure}
\centerline{
\includegraphics[]{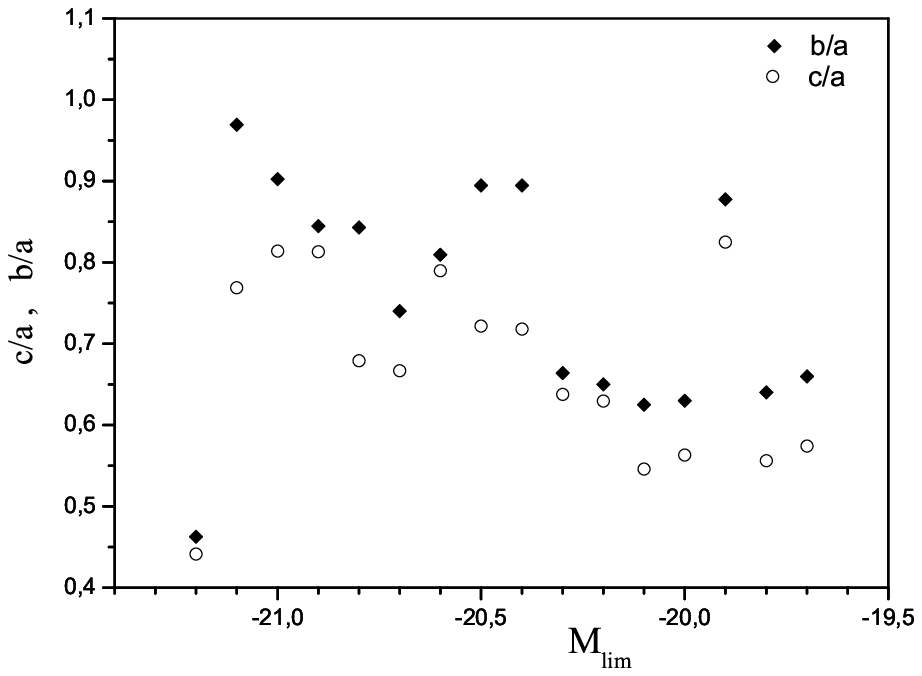}
} \figcaption{c/a and b/a ratios of one first void versus
$M_{lim}$.}
\end{figure}

\begin{figure}
\centerline{
\includegraphics[]{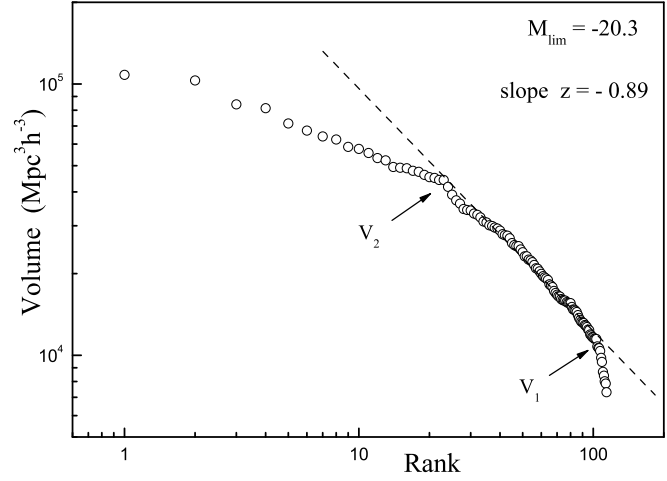}
} \figcaption{Void volume versus void rank at $M_{lim} = -20.3$.
The power-law segment in the volume interval $V_1 - V_2$.}
\end{figure}

\begin{figure}
\centerline{
\includegraphics[]{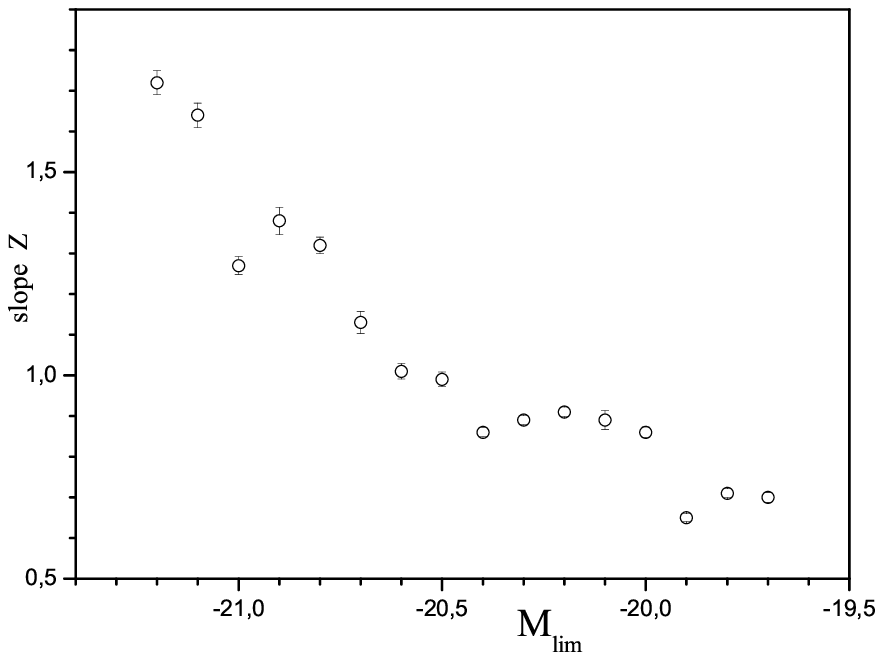}
} \figcaption{Evolution of the exponent of the "void volume ---
rank" relation in the characteristic volume interval with
$M_{lim}$.}
\end{figure}

\section{THE DISTRIBUTION OF VOID CENTERS}

In this section, we analyzed a sample with $z_{min} = 0.02$,
$z_{max} = 0.12$, $M^0_{lim} = -20.11$, in which 327 voids with
$R_{seed}>9h^{-1}$ Mpc were identified with a resolution of about
1.5 Mpc. The void centers were defined as the centers of mass of
the set of grid points assigned to a given void. We analyzed the
distribution of void centers using an cumulative correlation gamma
function (conditional density) (Coleman and Pietronero 1992;
Tikhonov 2006a). Once the signal has been stabilized (on small
scales, the signal is absent, because the distances between the
void centers are larger than 18 Mpc), the void centers exhibit a
correlation on scales $35 - 70h^{-1}$~Mpc with exponent $\gamma_v
\sim 0.5$ (Fig. 13). The galaxies involved in determining the
voids ($M_{abs} < 20.11$, except for isolated galaxies with
$R_{n}+\sigma_n \approx 3.5$~Mpc) show a slightly weaker
correlation, $\gamma_g \sim 0.4$, on these scales. These scales
correspond to the range of scales in the galaxy distribution in
which the transition to uniformity occurs (Tikhonov 2006a). The
slightly larger exponent for the void centers is apparently
obtained due to the empty regions near the boundaries, where there
are no void centers (according to the definitions in the void
search algorithm) and due to the differences in void volumes.

\begin{figure}
\centerline{
\includegraphics[]{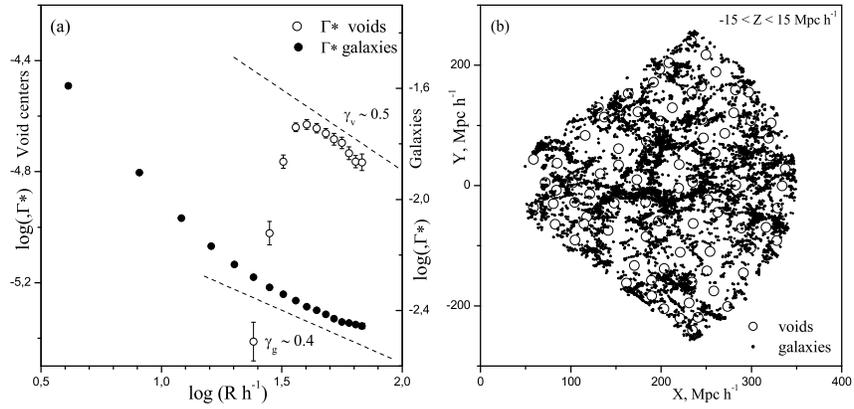}
} \figcaption{(a) Correlation functions of the galaxies (filled
circles) and void centers (open circles). The right and left Y
axes show the amplitudes of the correlation functions of the void
centers and galaxies, respectively. (b) The distribution of void
centers and galaxies in a $\pm15$~Mpc layer along the Z-axis (in
$\lambda$, $\eta$ coordinates).}
\end{figure}

\section{COMPARISON OF THE PROPERTIES
OF GALAXIES IN VOIDS AND STRUCTURES}

To compare the properties of the galaxies in voids and structures,
we constructed a volume-limited sample with $z_{min} = 0.02$ and
$z_{max} = 0.12$ with a limit on the absolute magnitude $M^0_{lim}
= -20.11$. The voids were determined in the distribution of
galaxies with $M_{abs} < M_{lim} = -20.44$ in such a way that some
the galaxies from the volume-limited sample could fall into voids.
We then excluded isolated galaxies (without any neighbors at a
distance smaller than $R_{n}+\sigma_n \approx 4.2$~Mpc) from the
resulting list of void-forming galaxies. In this way, a certain
number of isolated galaxies with $M_{abs} < -20.44$ fall into
voids.

We identified 235 voids with $R_{seed} > 10 h^{-1}$~Mpc. The voids
were divided into "large" ones with $R_{eff} = (3 \cdot Vol/4
\cdot\pi)^{1/3} > 20h^{-1}$~Mpc and "small" ones with $R_{eff} <
20h^{-1}$ Mpc. $R_{seed}$ and $R_{eff}$ are correlated: $R_{eff} =
(1.26\pm0.02) \cdot R_{seed}$. However, the void with a smaller
$R_{seed}$ (i.e., identified later) may have a larger volume. The
galaxies that fell into voids were divided into "bright" galaxies
of the volume-limited sample with $M_{abs} < -20.11$, i.e., those
without any selection in the radial direction, and "faint" ones
with $M_{abs}> -20.11$. A total of 2480 "bright" and 6104 "faint"
galaxies fell into large voids. The mean overdensity of the
galaxies from the volume-limited sample with $M_{abs} < -20.11$
that fell into voids is $\delta\rho/<\rho_{VL}> = -0.78$,where
$<\rho_{VL}>$ is the mean density of the galaxies with $M_{abs} <
M_{lim} = -20.11$.

The galaxy distribution in the identified voids shows the same
features as the galaxy distribution in the entire volume (as was
pointed out by Patiri et al., 2006b). Thus, for example, the
galaxy distribution in void ¹ 2 in order of identification in a
20Mpc layer in Z (in $\lambda$, $\eta$ coordinates) is essentially
nonuniform (Fig. 14) --- groups of galaxies and a filament
crossing the void can be identified. The galaxies delineate the
void boundaries and avoid the central region.

To compare the galaxy properties, we selected a check sample
containing galaxies in structures (dense regions). The galaxies in
structures, i.e., those that do not fall into voids, were
identified by constructing the minimum spanning tree. This tree
consists of knots and edges and is constructed by appending new
knots satisfying the condition for the distance to the already
constructed part of the tree being at a minimum (Barrow et al.
1985). We constructed the minimum spanning tree in a sample with
the angular boundaries $-30^\circ < \lambda < 30^\circ$, $10^\circ
< \eta < 30^\circ$.

Once the minimum spanning tree has been constructed from "bright"
galaxies with $M_{abs} < -20.11$, we identified the galaxies that
were connected by the edges of lengths no larger than $L_{max}$
determined from the assumed limit on the overdensity. The relation
between these parameters is defined by formula $\rho = 1/V =
3/4\cdot\pi\cdot L_{max}^3$. As the lower limit, we chose the
overdensity $\delta\rho/<\rho_{VL}> = 2$ ($<\rho_{VL}> \approx 4.5
\cdot 10^3$~Mpc$^{-3}$), which corresponds to an edge of length
$L_{max}=2.6h^{-1}$~Mpc. The mean distance to the nearest neighbor
in this sample is $R_n \approx 2.0h^{-1}$~Mpc. When the edges
smaller than $L_{max}$ are excluded from the tree, the tree breaks
up into connected "islands" with $\delta\rho/<\rho_{VL}> > 2$. For
our analysis, we retained only those "islands" that contained more
than 40 "bright" galaxies. In the sample under consideration,
these structures contain a total of 3011 "bright" galaxies. The
"faint" galaxies with $M_{abs} > -20.11$ offset by less than
$L_{max}$ from the island galaxies were then appended to these
structures. The number of "faint" galaxies attributed to
structures turned out to be 3533.

\newpage
\begin{figure}
\centerline{
\includegraphics[]{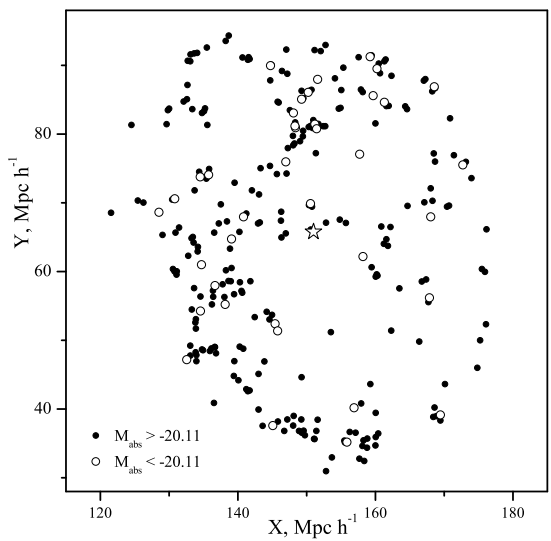}
} \figcaption{Distribution of galaxies in a $\pm10h^{-1}$~Mpc
layer along the Z-axis relative to the center of void No. 2 in
order of identification. The asterisk indicates the void center.}
\end{figure}

The distribution of the "faint" galaxies that fell into voids and
structures is subject to selection (incompleteness) in the radial
direction, but we will consider their properties separately from
those of the galaxies from the volume-limited sample --- the
influence of incompleteness is averaged to some degree, because
the voids and structures are distributed quite uniformly over the
sample volume.

Figure 15 presents the luminosity distribution for the galaxies in
voids and structures. On average, the galaxies in structures are
more luminous than those in voids --- an excess of "bright"
galaxies in structures and "faint" galaxies in voids is observed.
The abrupt cutoff of the histogram for the galaxies in voids near
$M_{abs} = -20.44$ results from the fact that only isolated
galaxies with $M_{abs} < -20.44$ fall into voids according to the
construction.

\begin{figure}
\centerline{
\includegraphics[]{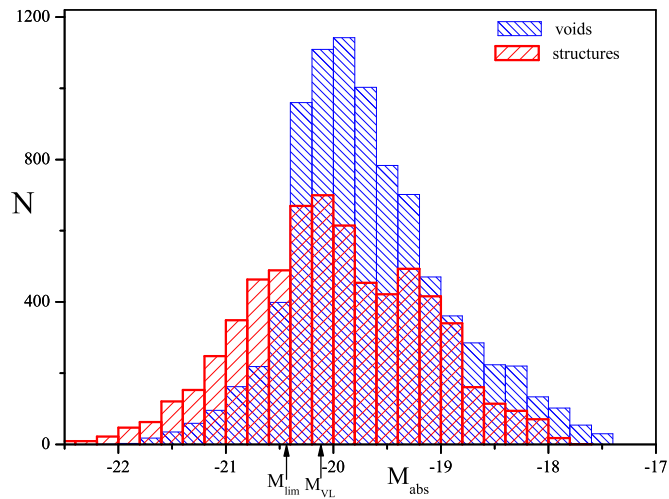}
} \figcaption{Distribution of galaxies in voids and structures in
absolute magnitudes (opposite hatching). $M_{VL}=-20.11$ is the
boundary of the volume-limited sample and $M_{lim}=-20.44$ is the
upper limit on the absolute magnitudes of the involved in
constructing the voids.}
\end{figure}

Figure 16 presents the galaxy number distribution in large voids
as a function of the galaxy distance $r$ from the void boundary
normalized to the effective void radius $R_{eff}$ for "bright"
galaxies with $-20.44 < M_{abs} < -20.11$ and isolated galaxies
with $M_{abs} < -20.44$ that fell into voids. Both subsamples show
a similar increase in the number of galaxies toward the void
boundaries. This, in particular, suggests that our approach to
excluding isolated galaxies from the list of galaxies in the
distribution of which we identified voids is legitimate.

\begin{figure}
\centerline{
\includegraphics[]{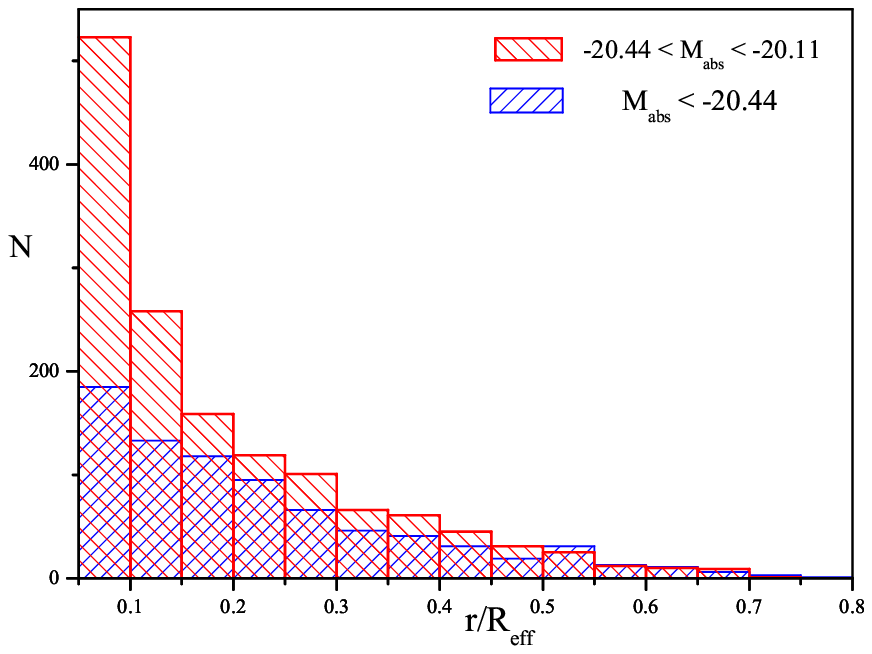}
} \figcaption{Galaxy number distribution as a function of the
distance from the void boundary normalized to the effective void
radius, $R_{eff}$, for all voids with $R_{eff} > 20h^{-1}$~Mpc.
Opposite hatching is used to denote the histograms for galaxies
with $-20.44<M_{abs}<-20.11$ and isolated galaxies in these voids
with $M_{abs}<-20.44$.}
\end{figure}

The galaxies in the identified voids exhibit a binomial (red and
blue) color distribution (Fig. 17). The mean color for "bright"
galaxies in large voids $<u-r>^v_{VL} = 2.22$,
$\sigma^v_{VL}=0.50$. For "faint" galaxies, $<u-r>^v_{dim} =
1.94$, $\sigma^v_{dim}=0.60$, The systematic difference reflects
the fact that the "faint" galaxies are, on average, bluer. Similar
characteristics are observed for the galaxies in small voids. For
"bright" and "faint" galaxies in structures, $<u-r>^s_{VL} =
2.47$, $\sigma^s_{VL}=0.50$ è $<u-r>^s_{dim} = 2.18$,
$\sigma^s_{dim}=0.57$. They are systematically redder than those
in voids. The histograms for the galaxies in voids and structures
differ significantly in shape --- the fraction of blue galaxies is
much larger in voids.

\begin{figure}
\centerline{
\includegraphics[]{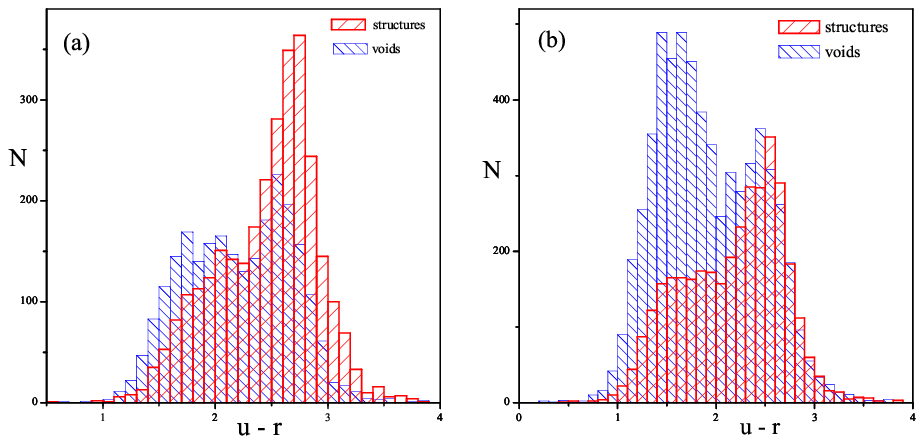}
} \figcaption{$u - r$ colors of galaxies in voids and structures
(the histograms with opposite hatching): (a) "bright" galaxies
with $M_{abs} < -20.11$, (b) "faint" galaxies with $M_{abs} >
-20.11$.}
\end{figure}

The data on the star formation rates per unit stellar mass for
SDSS galaxies ($log(SFR/M_{star}$), below referred to as SFR)
(Kauffmann 2003) were taken from the SDSS archive

\begin{figure}
\centerline{
\includegraphics[]{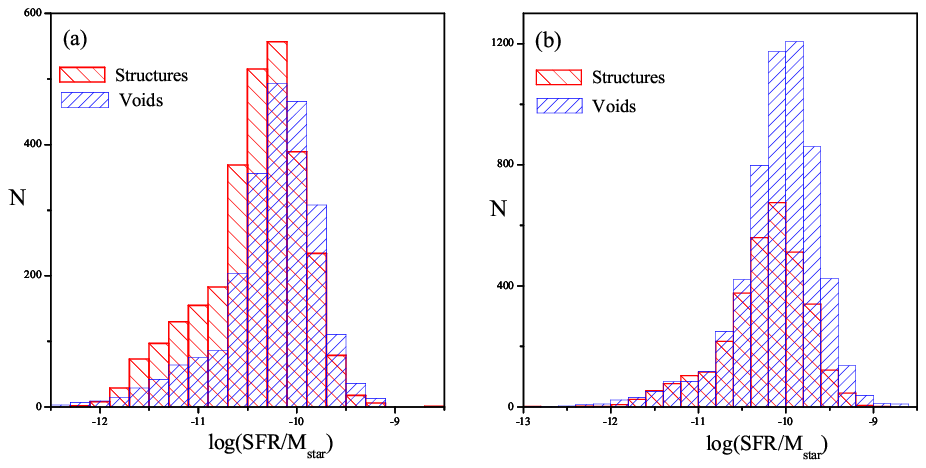}
} \figcaption{Star formation rates per unit stellar mass
($log(SFR/M_{star})$) for (a) "bright" and (b) "faint" galaxies in
voids and structures.}
\end{figure}

(http://www.mpagarching.mpg.de/SDSS/Dr4/). Figure 18 compares
$log(SFR/M_{star})$ for the galaxies in large voids and
structures. The mean value and dispersion of $log(SFR/M_{star})$
for large voids for "bright" galaxies are$<SFR>^v_{VL} = -10.26$,
$\sigma^v_{VL}=0.49$, respectively. For "faint" galaxies,
$<SFR>^v_{dim} = -10.09$, $\sigma^v_{dim}=0.49$. Again the
difference in the mean values stems from the fact that faint
galaxies have, on average, higher star formation rates than bright
galaxies. The "bright" and "faint" galaxies in structures have
$<SFR>^s_{VL} = -10.43$, $\sigma^s_{VL}=0.50$, $<SFR>^s_{dim} =
-10.25$, $\sigma^s_{dim}=0.49$, respectively. The star formation
rates in structures are systematically lower.

The red ($u-r > 2$) and blue ($u-r < 2$) galaxies show the
following characteristics for "bright" galaxies in large voids:
$<u-r>^v_{blue} = 1.68$, $\sigma^v_{blue}=0.21$; $<u-r>^v_{red} =
2.52$, $\sigma^v_{red}=0.42$ and in structures: $<u-r>^s_{blue} =
1.73$, $\sigma^s_{blue}=0.20$; $<u-r>^s_{red} = 2.63$,
$\sigma^s_{red}=0.37$. For "faint" galaxies in voids:
$<u-r>^v_{blue} = 1.56$, $\sigma^v_{blue}=0.26$; $<u-r>^v_{red} =
2.46$, $\sigma^v_{red}=0.39$ and in structures: $<u-r>^s_{blue} =
1.60$, $\sigma^s_{blue}=0.26$; $<u-r>^s_{red} = 2.51$,
$\sigma^s_{red}=0.41$.

The differences for voids and structures are found to be
insignificant but larger than those obtained in a similar analysis
by Patiri et al. (2006b).

For the galaxies in structures, the "($u-r$) color --- absolute
magnitude" (the trend with a slope of $slope_1 = -0.24$
--- brighter galaxies are, on average, redder),
"$log(SFR/M_{star})$ --- absolute magnitude ($slope_2 = 0.14$ ---
brighter galaxies have, on average, lower star formation rates),
and "color --- log(SFR/Mstar)" ($slope_3 = -0.59$ --- bluer
galaxies have, on average, higher star formation rates) relations
are less pronounced than those for the galaxies in voids, for
which $slope_1 = -0.27$, $slope_2 = 0.16$, $slope_3 = -0.66$,
respectively, although the differences are small.

The mean luminosity of the galaxies in all the identified voids is
virtually independent of the galaxy distance from the void
boundary normalized to the effective void radius $R_{eff}$. The
"bright" and "faint" galaxies in voids show opposite weak trends:
on average, the galaxies with $M_{abs} < -20.11$ become brighter
and the galaxies with $M_{abs} > -20.11$ become fainter as one
goes from the boundaries to the central regions of the voids.

The ("bright" and "faint") galaxies in voids, on average, have
bluer colors (lower $u-r$) and higher star formation rates (the
slopes of the linear fits to the "(($u-r$) --- $r/R_{eff}$ è
$log(SFR/M_{star})$ --- $r/R_{eff}$" relations are -0.12 and 0.06,
respectively) as $r/R_{eff}$ increases (where $r$ is the distance
from the boundary of a given void). Thus, the differences in the
properties of the galaxies in voids and structures slightly
increase even further when less dense regions inside the
identified voids are considered.

\section{CONCLUSIONS}

In this paper, using an algorithm of identifying arbitrarily
shaped voids, we analyzed the change in some of the void
characteristics with luminosity of the galaxies (in the range of
limits $M_{lim}$ on the absolute magnitude -19.7 --- -21.2)
involved in the construction of voids and compared the properties
of the galaxies in voids (mean density contrast) and the galaxies
forming structures with an density contrast higher than 2.

The evolution of a number of void characteristics with luminosity
shows a break near M*, in agreement with the change in exponent in
a correlation analysis of the distribution of SDSS galaxies
(Tikhonov 2006a). The structure of the galaxy distribution
apparently changes qualitatively at $M_{abs} < M^*$.

As expected, the void volumes increase with decreasing $M_{lim}$
with a break near M* followed by a faster increase (in contrast,
the decrease in the number of galaxies that form voids slows
down), with the pattern being the same for two samples of equal
volume covering different regions of space.

The mean density contrast in voids also increases with decreasing
$M_{lim}$ with a weak break near M*. The voids become "emptier" if
we reduce the limit on the absolute magnitude of the galaxies
under consideration at fixed $M_{lim}$.

The galaxies inside voids concentrate to the void boundaries and
avoid the central regions. The density contrast profile is flat in
intermediate regions. These results agree with those obtained
previously from the galaxies (Patiri et al. 2006b) and the dark
matter haloes in voids (Gottlober et al. 2003). Our study also
confirms that the matter in voids is distributed irregularly and
has the same features as the general distribution obtained by the
above authors.

In general, the mean characteristics of the void shapes are
retained with decreasing $M_{lim}$. We can only note a weak
tendency: at $M_{lim}$ < -20.6, the voids become, on average,
slightly more elongated. There is also a weak tendency for the
slope of the distributions of the medium-to-larger (b/a)
smaller-to-larger (c/a) axial ratios of the ellipsoid equivalent
to the void (these ratios are well correlated) to increase with
decreasing $M_{lim}$. The slopes of the linear fits to these
distributions are close to unity (0.8 -- 1.1) in the entire
$M_{lim}$ range, i.e., the voids are predominantly elongated and
nonoblate at all $M_{lim}$. At the same time, the individually
considered voids can change their shape with $M_{lim}$
significantly. The directions of the greatest void elongations are
distributed quite uniformly and change chaotically with $M_{lim}$,
which is indicative of an irregular change in the structure when
more luminous galaxies are considered.

The exponent of the "void volume -- rank" ($log(Vol)$ --
$log(Rank)$) relation in the characteristic range of volumes
increases significantly with decreasing $M_{lim}$ starting from
$M_{lim} = -20.4$ (a break is again observed in the relation at a
value close to M*), which is a reflection of the tendency for more
luminous galaxies to cluster more strongly. The scales of the
beginning of the transition to uniformity (the break of the
power-law segment in the $Vol$ -- $Rank$ relation at volume $V_2$
-- see the text) determined from this relation increase with
decreasing $M_{lim}$ and agree with the results of a correlation
analysis of the galaxy distribution.

The distribution of void centers shows a certain correlation and
reflects the correlations of the galaxy distribution on the
corresponding scales.

The derived differences in the properties of the galaxies in voids
with a mean density contrast of -0.78 and the galaxies in
structures with an density contrast higher than 2 (identified
using the minimum spanning tree) agree qualitatively with the
existing views: the galaxies in voids are, on average, bluer and
have higher star formation rates per unit stellar mass. The last
two tendencies become stronger when galaxies located closer to the
central void regions are considered. However, quantitatively,
these differences are not large enough to conclude that the
formation histories of the galaxies in structures and voids differ
fundamentally.

The need for a further study of the observed correlations and
multiparameter tendencies and for determining the mean galaxy
characteristics over wide ranges of density contrasts and
luminosities and for various definitions of the characteristic
galaxy environments is obvious. Our division into voids and
structures is to some extent arbitrary and consists mainly in
different density contrasts: there are also structures in the
voids determined in this paper.

\section{ACKNOWLEDGMENTS}
This work was supported by grant No. MK-6899.2006.2 of the
President of Russian Federation.

\section{REFERENCES}

1. J. K. Adelman-McCarthy, M.A.Agueros, S.S.Allam, et al.,
Astron.J.(in press).

2. I. Baldry, K. Glazebrook, and J. Brinkmann, Astrophys. J. 600,
681 (2004).

3. J. Barrow, S. Bhavsar, and D. Sonoda, Mon. Not. R. Astron. Soc.
216, 17 (1985).

4. A. J. Benson, F. Hoyle, F. Torres, and M. Vogeley, Mon. Not. R.
Astron. Soc. 340, 160 (2003).

5. M. Blanton, J. Brinkmann, and I. Csabai, Astron. J. 125, 2348
(2003).

6. J. Colberg, R. Sheth, A. Diaferio, et al., Mon. Not. R. Astron.
Soc. 360, 216 (2005).

7. P. H. Coleman and L. Pietronero, Phys. Rep. 213, 311 (1992).

8. D. J. Croton, M. Colles, E. Gaztanaga, et al., Mon. Not. R.
Astron. Soc. 352, 828 (2004); astro-ph/0401406.

9. H. El-Ad and T. Piran, Astrophys. J. 491, 421 (1997).

10. S. Furlanetto and T.Piran, Mon.Not. R.Astron. Soc. 366, 467
(2006).

11. S. Furlanetto and T.Piran, Mon.Not. R.Astron. Soc. 366, 467
(2006).

12. J. Gaite, Europ. Phys.J.B 47, 93 (2005).

13. J. Gaite and S.C.Manrubia, Mon. Not. R. Astron. Soc. 335, 977
(2002); astro-ph/0205188.

14. S.Gottlober, E.L.Locas, A.Klypin, and Y.Hoffman, Mon. Not. R.
Astron. Soc. 344, 715 (2003); astro-ph/0305393.

15. D.W. Hogg, astro-ph/9905116 (1999).

16. D. Hogg, M. Blanton, J. Brinchmann, et al., Astrophys. J. 601,
L29 (2004).

17. F. Hoyle and M. S. Vogeley, Astrophys. J. 607, 751 (2004);
astro-ph/0312533.

18. F. Hoyle, R. Rojas, M. Vogeley, and J. Brinkmann, Astrophys.
J. 620, 618 (2005).

19. B. V. Icke, Mon.Not.R.Astron. Soc. 206, 1 (1984).

20. G. Kauffmann, T. Heckman, S. White, et al., Mon. Not. R.
Astron. Soc. 341, 33 (2003).

21. S. G. Patiri, J. Betancort-Rijo, F. Prada, et al., Mon. Not.
R. Astron. Soc. 369, 335 (2006a); astro-ph/0506668.

22. S. Patiri, F. Prada, J. Holtzman, et al., Mon. Not. R. Astron.
Soc. 372, 1710 (2006á).

23. P. J. E. Peebles, Astrophys. J. 557, 495 (2001).

24. M. Plionis and S. Basilakos, Mon. Not. R. Astron. Soc. 330,
399 (2002).

25. E. Regoes and M.Geller, Astrophys.J. 377,14 (1991).

26. R. Rojas, M. Vogeley, F. Hoyle, and J. Brinkmann, Astrophys.
J. 617, 50 (2004); astro-ph/0307274.

27. R. Rojas, M. Vogeley, F. Hoyle, and J. Brinkmann, Astrophys.
J. 624, 571 (2005).

28. W. Saunders, W. J. Sutherland, S. J. Maddox, et al., Mon. Not.
R. Astron. Soc. 317, 55 (2000).

29. S. Shandarin, J. Sheth, and V. Sahni, Mon. Not. R. Astron.
Soc. 353, 162 (2004).

30. S. Shandarin, H. A. Feldman, K. Heitmann, and S.Habib, Mon.
Not.R.Astron. Soc. 367, 1629 (2006).

31. R. Sheth and R. van de Weygaert, Mon. Not. R. Astron. Soc.
350, 517 (2004).

32  C. Stoughton, R. Lupton, M. Bernardi, et al., Astron. J. 123,
485 (2002).

33. A. V. Tikhonov, Pis'ma Astron. Zh. 32, 803 (2006a) [Astron.
Lett. 32, 721 (2006a)]; astro-ph/0610643.

34. A. V. Tikhonov, Pis'ma Astron. Zh. 33, 809 (2006b) [Astron.
Lett. 33, 727 (2006b)]; astro-ph/0610689.

35. A. V. Tikhonov and I. D. Karachentsev, Astrophys. J. 653, 969
(2006).

\end{document}